\documentclass[12pt]{article}
\pdfoutput=1
\usepackage{amsmath}
\usepackage{amsfonts}
\usepackage{setspace}
\usepackage{subcaption}
\usepackage{enumerate}
\usepackage{graphicx}
\usepackage[hidelinks]{hyperref} 
\numberwithin{equation}{section}
\usepackage[utf8]{inputenc}
\newcommand{\gsim}{\lower.7ex\hbox{$\;\stackrel{\textstyle>}{\sim}\;$}}
\newcommand{\lsim}{\lower.7ex\hbox{$\;\stackrel{\textstyle<}{\sim}\;$}}
\def\O{{\mathcal O}}

\newcommand{\be}{\begin{equation}}
\newcommand{\ee}{\end{equation}}
\newcommand{\bea}{\begin{eqnarray}}
\newcommand{\eea}{\end{eqnarray}}

\newcommand{\comment}[1]{}

\newcommand{\Cint}{C\kern-1em\int}

\def\d{\partial}
\def\vphi{\varphi}

\def\O{\mathcal{O}}

\def\ad{{\rm AdS}_2}
\def\ds{{\rm dS}_2}
\begin{document}
\vspace*{-1. cm}
\begin{center}
{\bf \Large Uptunneling to de Sitter}
\vskip 1cm
{{\bf Mehrdad Mirbabayi} }
\vskip 0.5cm
{\normalsize {\em International Centre for Theoretical Physics, Trieste, Italy}}
\vskip 0.2cm
{\normalsize {\em Stanford Institute for Theoretical Physics, Stanford University,\\ Stanford, CA 94305, USA}}
\end{center}
\vspace{.8cm}
{\noindent \textbf{Abstract:}  
We propose a Euclidean preparation of an asymptotically AdS$_2$ spacetime that contains an inflating dS$_2$ bubble. The setup can be embedded in a four dimensional theory with a Minkowski vacuum and a false vacuum. AdS$_2$ times 2-sphere approximate the near horizon geometry of a $4d$ near-extremal RN wormhole. 
Likewise, in the false vacuum the near-horizon geometry of a near-extremal black hole is approximately dS$_2$ times 2-sphere. We interpret the Euclidean solution as describing the decay of an excitation inside the wormhole to a false vacuum bubble. The result is an inflating region inside a non-traversable asymptotically Minkowski wormhole. 
\vspace{0.3cm}
\vspace{-1cm}
\vskip 1cm
\section{Introduction}
False vacua are accessible in classical and quantum field theory, given enough time and energy. By a careful preparation of the initial state, an experimentalist could, in principle, create an arbitrarily big region of false vacuum in the laboratory. For instance, given that any finite region of the false vacuum will fully decay into asymptotic QFT states, a recipe would be to apply time-reversal to the decay products of a state that contains such a region to begin with. 

In gravity, the situation is qualitatively different. A big region of a false vacuum with positive cosmological constant (CC) can eternally inflate, and as a result change the asymptotic structure of the spacetime. In fact, in classical gravity Penrose singularity theorem \cite{Penrose} forbids the formation of such an inflating bubble in a Minkowski vacuum with no singularity in the past \cite{Guth}. It is tantalizing to ask whether this can happen quantum mechanically. It would be quite remarkable if quantum gravity completely excised macroscopic domains of de Sitter vacua from Minkowski and Anti-de Sitter physics. 

But if it didn't, one could hope to use the false vacuum bubble to shed some light on the notoriously hard to understand quantum mechanics of de Sitter (dS) spacetime. We have a better handle on quantum gravity in asymptotically Minkowski spacetime via scattering amplitudes and in asymptotically Anti-de Sitter (AdS) spacetime via holography. If dS arises in the excitations of the Minkowski or AdS vacua, those frameworks can, at least in principle, be employed to also explore dS quantum gravity. 

On the other hand, dS can help formulate new problems in the asymptotically Minkowski or AdS setups. For instance, because of the negative interior pressure, dS bubbles are generically hidden behind black hole horizons. This provides an interesting interplay between black hole microstate counting and that of de Sitter.\footnote{It is actually known that in an FRW universe with inflation (a period of quasi-dS expansion) in its past there can be black holes connecting to inflating pocket universes \cite{Garriga}. Nevertheless, it would be desirable if we could study dS without starting from it.}

But perhaps the best motivation is coming from the cosmological observations. They suggest that we live in a dS false vacuum with a tiny CC, and that our universe has gone through a period of inflation at very early times. One might wonder: Could it all have started inside the laboratory of a dedicated experimentalist?

This question was raised 30 years ago in \cite{Farhi,Fischler,Fischler2}. To go around the singularity theorem, they invoke quantum tunneling to a spacetime that if classically extended it would contain a singularity in the past. However, the solution of \cite{Farhi,Fischler,Fischler2} is degenerate, and to date there is no consensus on whether or not it describes a valid tunneling process. See \cite{Quevedo} and \cite{Fu} for two recent papers with contrasting views and for further references. To break the impasse, finding an alternative solution seems necessary. 

Here we propose an alternative. One that is a byproduct of our solution (in $2d$ gravity) to a similarly interesting question raised in \cite{Fu}: Is it possible to prepare, via a Euclidean path integral, an asymptotically AdS state that contains an inflating bubble?

The two questions are connected if we take advantage of simplifications that arise in the near horizon geometry of near-extremal black holes. Specifically, our focus is on the near extremal Reissner-Nordstr\"om (RN) black hole in the Minkowski vacuum and on the near extremal Schwarzschild-de Sitter (SdS) black hole in the false vacuum. Both geometries have a long throat with an approximately constant radius and hence one can dimensionally reduce to obtain a $2d$ gravity model called Jackiw-Teitelboim (JT) gravity \cite{Jackiw,Teitelboim}. The $2d$ geometry is AdS$_2$ in the RN case and dS$_2$ in the SdS case. These connections have been the subject of extensive research in recent years. Two papers that have been particularly inspiring to us are \cite{Maldacena_AdS,Maldacena_dS}, though for completeness a brief review will be given in section \ref{sec:JT}.

In section \ref{sec:sol}, we find a Euclidean solution in which a brane emanates from the AdS$_2$ boundary, it decays into a dS$_2$ bubble (a piece of a 2-sphere in Euclidean signature), and bounces back to reach the AdS boundary (see figure \ref{fig:bounce}). Cutting this bounce solution at the moment of time-reflection symmetry and Wick rotating gives the desired asymptotically AdS$_2$ geometry with an inflating bubble in the middle (see figure \ref{fig:penrose}). Once embedded in four dimensions, the two AdS asymptotics match to the mouths of a near-extremal RN wormhole. The dS$_2$ region describes the near-extremal SdS geometry. And the domain walls carry opposite magnetic charges to break the magnetic field lines.

\begin{figure}[t]
\centering
\includegraphics[scale =0.8]{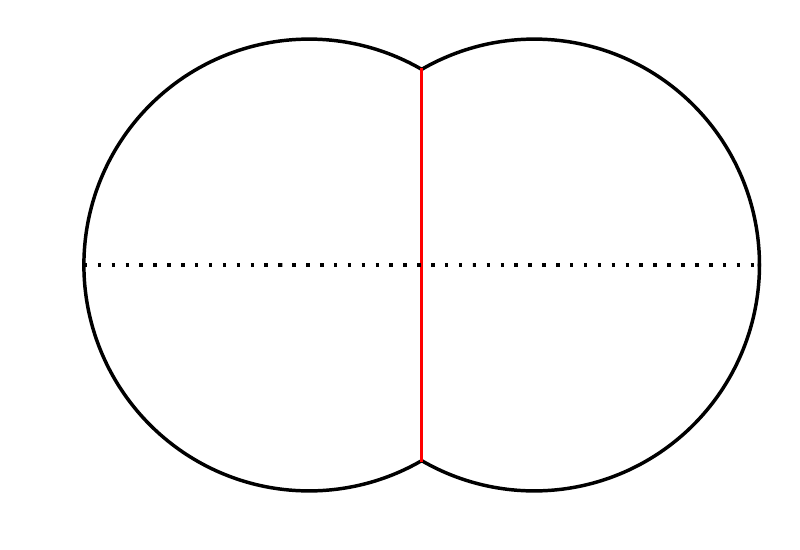} 
\includegraphics[scale =0.8]{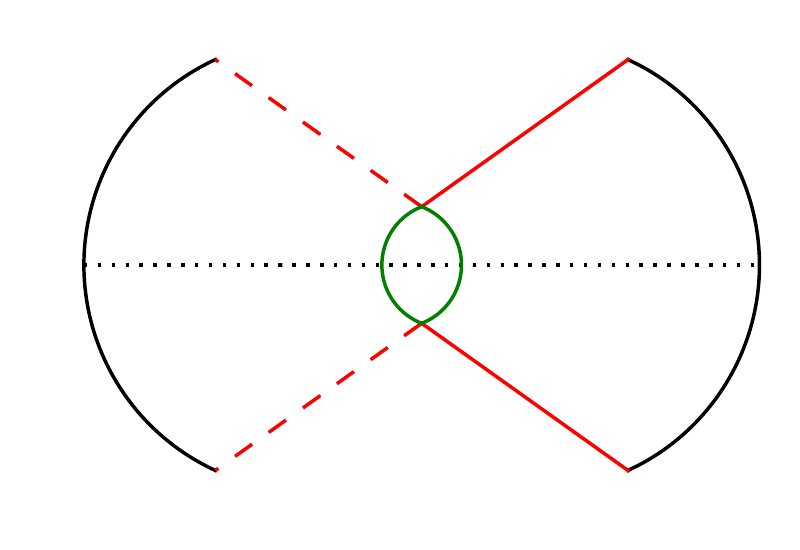} 
\caption{\small{{\em Schematic}. Left: a freely propagating brane in Euclidean AdS$_2$. Right: the bounce solution for the decay of the brane into a dS bubble. The dashed red lines on the left have to be identified with the solid red lines on the right. The diagrams can be cut at the moment of time-reflection symmetry (along the dotted lines) and Wick rotated to Lorentzian signature.}}
\label{fig:bounce}
\end{figure}
\begin{figure}[t]
\centering
\includegraphics[scale =0.8]{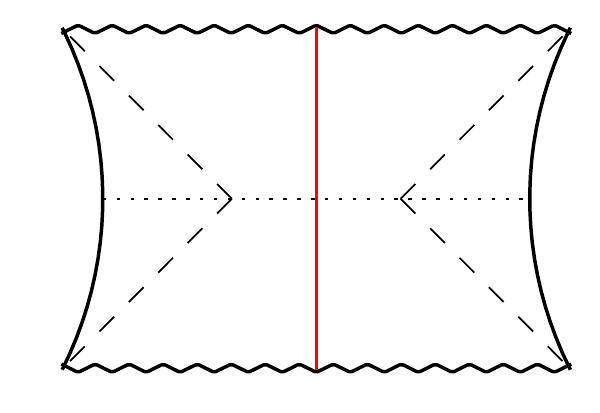} 
\includegraphics[scale =0.85]{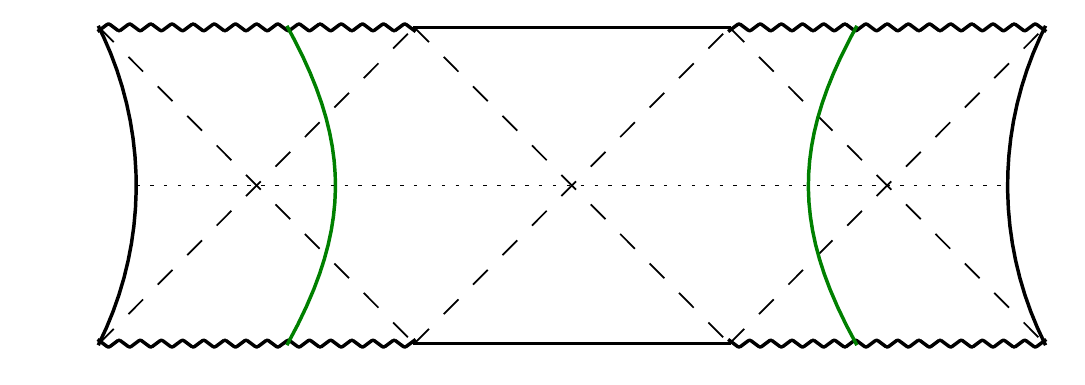} 
\caption{\small{Penrose diagrams obtained from the Wick rotation of the Euclidean solutions in figure \ref{fig:bounce} and analytic continuation to the past. Left: A two sided AdS$_2$ black hole with an elongated throat due to the presence of the massive brane. Right: An inflating bubble that is nucleated inside the AdS$_2$ throat. The domain walls (green) fall though the black hole horizons. In the $4d$ embedding, the inflating region has the spatial topology of $\mathbf{R}\times S^2$. At first, the $\mathbf{R}$ factor expands exponentially and the 2-sphere slowly. Eventually, the expansion becomes isotropic and we obtain a locally dS$_4$ solution. As seen, the classically extended geometry is singular in the past as dictated by the Penrose singularity theorem \cite{Penrose,Vilenkin}.}}
\label{fig:penrose}
\end{figure}
In section \ref{sec:con}, we will speculate about the tunneling probability by comparing the bounce action to the Euclidean action for the free propagation of the brane in AdS$_2$. We will conclude by further remarks on the viability of this scenario.
\section{JT gravity}\label{sec:JT}
Our goal is to study the motion of codimension-1 domain walls and branes in the RN and SdS geometries. Since we are mainly concerned with the Euclidean solution, it is enough to look at the static patch metric:
\be
ds^2 = -f(r) dt^2 + \frac{dr^2}{f(r)} + r^2 d\Omega^2
\ee
where $d\Omega^2$ is the line element on unit 2-sphere $S^2$, and 
\be\label{fS}
f(r) = 1-\frac{8\pi G}{3} \Lambda r^2 -\frac{2G M_i}{r}, \qquad \text{SdS}
\ee
where $\Lambda$ is the vacuum energy of the false minimum, and 
\be\label{fR}
f(r)=1-\frac{2 GM_e}{r} + \frac{4\pi G Q^2}{r^2},\qquad\text{RN}.
\ee
The horizons correspond to the zeros of $f(r)$. The extremal limit is when the two zeroes coincide. In the SdS case this is at
\be
r_0^2 = \frac{1}{8\pi G \Lambda}.
\ee
Suppose we tune the masses $M_i$, $M_e$ and the magnetic charge $Q$ such that both geometries are near extremality and with approximately equal horizon areas. Then the near horizon geometry can be studied by dimensionally reducing over $S^2$ and working with the $2d$ model. Ignoring the KK modes and setting $r_0=1$, we have the spherically symmetric metric ansatz
\be
ds^2 = g_{\mu\nu} dx^\mu dx^\nu + (1+\phi) d\Omega^2,\qquad \phi \ll 1
\ee
where $\mu,\nu$ run over $0$ and $1$. Dropping topological terms, the $4d$ action reduces to $2d$ JT gravity coupled to $2d$ matter fields $\psi$:
\be\label{JT}
S = C \left[\int d^2x \sqrt{-g} (\phi R - U(\phi))-2\phi_b\oint d\ell\ k\right] + 4\pi S_m[g_{\mu\nu},\psi]
\ee
where in terms of the $4d$ Newton's constant $C =1/(4 G)$. $\phi$ is called the dilaton field. At $\phi =\phi_b\sim 1$ the $2d$ theory has to be matched with the higher dimensional one. $k$ is the geodesic curvature of this boundary, and $d\ell$ its line element. Having set the extremal radius to one, the dilaton potential is 
\be
U(\phi) = \left\{\begin{array}{cc} 2\phi,\qquad \text{SdS}\\[10pt] -2\phi,\qquad \text{RN}\end{array}\right.
\ee
To leading order in $\phi$ the $2d$ metric in the false vacuum is the $\ds$ metric with unit radius of curvature.\footnote{Earlier studies of $\ds$ physics include \cite{Anninos,Galante,Cotler,Maldacena_dS}. However, in some cases the possibility of embedding in a higher dimensional setup has not been a requirement.} Working in the static patch and Wick rotating, we get the sphere metric
\be\label{ds}
ds^2 = d\theta^2 +\sin^2\theta \ d\vphi^2.
\ee
This can be derived from the variation of \eqref{JT} with respect to $\phi$, or directly from \eqref{fS} by expanding in the near-horizon, near-extremal limit. The $\phi$ solution is given by
\be\label{phids}
\phi = B \cos\theta,\qquad B = 2\sqrt{\frac{2}{3}-2 G M_i}.
\ee
The points $\theta=0$ and $\theta=\pi$ correspond, respectively, to the cosmological and the black hole horizons of the SdS geometry. Beyond the cosmological horizon, $\phi$ expands and one eventually recovers isotropic $4d$ inflation. The form of the solution \eqref{phids} can also be derived by varying the effective $2d$ action \eqref{JT} with respect to $g_{\mu\nu}$, which gives
\be\label{phieq}
(g_{\mu\nu} \nabla^2 - \nabla_\mu\nabla_\nu)\phi+\frac{1}{2} g_{\mu\nu}U(\phi) = \frac{2\pi}{C} T_{\mu\nu},
\ee
and setting $T_{\mu\nu} =0$. Matter perturbations back-react on $\phi$, and in order for the approximation $\phi\ll 1$ to hold true, we need (restoring factors of $r_0$)
\be
G r_0^2 T_{\mu\nu}=\O(\phi) \ll 1.
\ee
We neglect the small effect of these perturbations on the curvature of the $2d$ metric $g_{\mu\nu}$.

In the RN case, the extremal radius of a charge $Q$ black hole is
\be
r_e^2 =4\pi G_N Q^2.
\ee
We take $r_e\simeq 1$. In the near extremal limit, the $2d$ metric is approximately a unit-curvature AdS$_2$. In Euclidean signature
\be\label{ads}
ds^2 = d\rho^2 + \sinh^2\rho \ d\vphi^2,
\ee
as follows from the variation of \eqref{JT} with respect to $\phi$, or directly expanding \eqref{fR} and Wick rotating. Allowing for a small offset between the extremal radii $r_e$ and $r_0 =1$, corresponds to including a small $2d$ CC on the r.h.s. of \eqref{phieq}. The $\phi$ solution is then
\be\label{phiads}
\phi = A \cosh \rho - B_0, \qquad B_0 = 1 - r_e^2,
\ee
and in terms of the RN mass parameter $M_e$
\be\label{A}
A = 2 r_e \sqrt{2 r_e(GM_e-r_e)}.
\ee
We neglect $\O(\phi)$ corrections to the $2d$ geometry. 
\section{Domain wall motion}\label{sec:sol}
The motion of a domain wall follows from the junction condition between the two geometries it connects, and it is particularly simple in the spherically symmetric case \cite{Israel,Blau}. In our approximation, it follows from \eqref{phieq}:
\be\label{junc}
\left.\xi^\mu \d_\mu \phi\right|_L^R = \kappa, 
\ee
where L/R label the two sides, $\xi^\mu$ is the normal to the domain wall trajectory, pointing from right to left, and $\kappa$ is related to the brane tension $\sigma$ via
\be
\kappa = 8\pi G \sigma.
\ee
Our bounce solution consists of a brane that emanates from the AdS$_2$ boundary and bifurcates into the dS-AdS domain walls as in figure \ref{fig:bounce}.\footnote{Of course, in $2d$ these are just particles, but we continue calling them by their $4d$ names.} Below we will discuss each part separately.
\subsection{Brane in AdS$_2$}
In the absence of the brane the metric and dilaton are given by \eqref{ads} and \eqref{phiads} respectively. Next we add a brane that stretches to the AdS boundary and impose $\mathbf{Z}_2$ symmetry across it. When the two sides of the brane are identified, this is often called an end of the world brane. It is considered as a model for a one-sided black hole formed from the collapse of a pure state \cite{Kourkoulou}. See also \cite{Cooper,Antonini} for possible connections to cosmology. We do not make this identification. The normal vector to the brane is 
\be
\xi^\mu = (\sqrt{1-\dot \rho^2},\frac{\dot \rho}{\sinh \rho})
\ee
where over-dot indicates derivative with respect to the proper length. From \eqref{junc}, we find
\be\label{dotr}
\dot \rho = \sqrt{1- \frac{\sinh^2 \rho_m}{\sinh^2 \rho}},\qquad \sinh \rho_m = \frac{\kappa_0}{2A},
\ee
where the normalized brane tension is called $\kappa_0$. This is the equation of a geodesic in AdS$_2$. We could also arrive at this conclusion by using the fact that in the presence of a soft source (the brane) the $2d$ curvature has to remain finite. Denoting by $\eta$ the normal coordinate to the brane, the scalar curvature near the brane is related to its geodesic curvature via
\be\label{kLR}
R = 2(k_L-k_R) \delta(\eta) + \rm{regular}.
\ee
The $\mathbf{Z}_2$ symmetry implies $k_R=-k_L$, hence we must have $k_L=k_R=0$. Strictly speaking, the curve has a small $\O(\phi)$ curvature that is irrelevant for our discussion. 

The angular position of the brane at radial coordinate $\rho$, as measured from the closest approach to the origin $\rho= \rho_m$, can be calculated using \eqref{dotr},
\be\label{cosphi}
\cos \vphi = \frac{\tanh \rho_m}{\tanh \rho}.
\ee
Finally, the brane meets the $\ad$ boundary at $\rho=\rho_c$ with an angle 
\be\label{alb}
\alpha_b = \arccos\sqrt{1-\frac{\sinh^2\rho_m}{\sinh^2\rho_c}},
\ee
with respect to the normal. This contributes to the boundary term in the JT action \eqref{JT} because $k=2 \alpha_b \delta(u)+$regular, where $u$ is the proper length along $\d\ad$ measured from this point.

Without any decay, this solution can be Wick rotated at $\rho = \rho_m$ to give a two-sided eternal black hole with a brane inside as in figure \ref{fig:penrose}-Left.
\subsection{Brane decay}\label{sec:decay}
Suppose at $\rho=\rho_1$ the brane branches into two domain walls separating Euclidean $\ad$ from a Euclidean $\ds$ (i.e. $S^2$) region. First, we discuss the geometric aspects of the decay. At the end of this subsection, we will comment on the microscopic aspects. 

Let us first show that the $2d$ geometry is smooth along the domain wall. This follows from the derivative of \eqref{junc} along the domain wall. On the right we get $u^\nu \d_\nu \kappa$ and on the left
\be
\left.u^\nu \nabla_\nu (\xi^\mu \d_\mu \phi)\right|^R_L
=(u^\nu \nabla_\nu \xi^\mu)\d_\mu\phi|_L^R + u^\nu \xi^\mu \nabla_\nu\nabla_\mu\phi|_L^R .
\ee
The first term can be written in terms of the geodesic curvature, and in the second, we can use \eqref{phieq} and the fact that $u\cdot \xi = 0$ to find
\be\label{lr}
(k_R-k_L)u^\mu \d_\mu\phi +8\pi G u_\nu \xi_\mu (T_L^{\mu\nu}-T_R^{\mu\nu})= u^\nu \d_\nu \kappa.
\ee
By energy-momentum conservation the second term on the left is the same as the term on the right. Therefore\footnote{In the JT framework, the same result could be obtained by introducing boundary terms on the two sides of the domain wall and imposing that $\phi$ and $g_{\mu\nu}$ are continuous.}
\be\label{klr}
k_R = k_L \Rightarrow k_{AdS} = k_{dS}.
\ee
This holds also at the branching point, where the brane can be smeared over some width and be thought of as a flux of Euclidean energy from the AdS side in \eqref{lr} that is absorbed by the domain wall. Equation \eqref{klr} implies that there is no conical singularity at that point and the sum of the three exterior angles is $2\pi$. Note that if the boundaries of the left AdS region, the right AdS region and the dS part were all smooth curves, the sum of the angles would be $3\pi$. Therefore, there has to be a break in the boundary trajectories (see figure \ref{fig:bifur}).
\begin{figure}[t]
\centering
\includegraphics[scale =0.8]{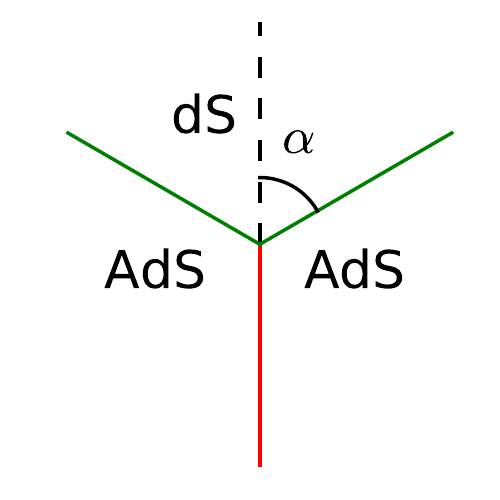} 
\caption{\small{The brane bifurcation into the dS-AdS domain walls.}}
\label{fig:bifur}
\end{figure}
Denote the break of the right AdS boundary by $\alpha$ and the $\rho$ velocity after the break $\dot \rho_+$, then
\be\label{alpha}
\sqrt{1-\frac{\sinh^2 \rho_m}{\sinh^2 \rho_1}} \dot \rho_+ + \frac{\sinh \rho_m}{\sinh \rho_1} \sqrt{1-\dot \rho_+^2}= \cos\alpha.
\ee
On the dS side, using $\mathbf{Z}_2$ symmetry and imposing that the bubble includes the cosmological horizon at $\theta=0$, we find
\be\label{alpha1}
\dot\theta = -\cos\alpha, \qquad \text{at bifurcation}
\ee
where $\dot \theta$ is measured in the direction away from the branching point. 

In addition, \eqref{klr} forbids the formation of a dS bubble that is carved out of AdS without a deficit angle because otherwise
\be
\oint d\ell \  k_{dS} <2\pi < \oint d\ell \ k_{AdS} \qquad \text{no angular deficit.}
\ee
Indeed, if the AdS region in the bounce solution of figure \ref{fig:bounce} is continued beyond the bubble walls it either encounters a piece of the AdS boundary or the brane trajectories collide at an angle, implying a conical deficit. 

Finally, momentum conservation relates $\alpha$, $\kappa_0$ and the domain wall tension $\kappa$:
\be\label{mom}
\kappa_0 = 2 \kappa \cos\alpha.
\ee
Hence $\alpha < \pi/2$, and $\dot\theta <0$ at the branching point.

So far, our treatment of the brane decay has been purely phenomenological. A possible microscopic realization is to consider the brane to be the fundamental particle (FP) describing scalar excitations around the true vacuum. There is a cubic coupling between FP and the domain walls (kink and anti-kink in the limit of degenerate vacua). At weak coupling, FP is much lighter than the domain walls and hence the decay process is kinematically forbidden. This corresponds to a decay process that is allowed in Euclidean signature, as in our phenomenological model (see \eqref{mom}). Our finding is that when gravity is included, the decay products can materialize in the Lorentzian geometry. This is somewhat analogous to the decay of a light axion into an $e^+e^-$ pair. If $m_a <2 m_e$, the decay is kinematically forbidden in vacuum but can happen in a strong electric field.


\subsection{dS-AdS domain wall}
The Euclidean dS$_2$ metric and dilaton solutions are given by \eqref{ds} and \eqref{phids}. The continuity of $\phi$ at the domain wall implies
\be\label{contin}
A\cosh \rho = B \cos \theta + B_0.
\ee
To simplify equations, we assume the bubble forms at $\rho_1\gg 1$, and therefore there is a hierarchy 
\be
A\ll B,\qquad A\ll B_0.
\ee
Then, we get from \eqref{contin}
\be\label{dotr1}
\dot \rho = - \frac{B\sin\theta}{B\cos\theta + B_0} \dot\theta.
\ee
The opposite sign of $\dot \rho$ and $\dot\theta$ is because we are keeping the cosmological horizon at $\theta=0$, rather than black hole horizon $\theta =\pi$. Therefore, larger values of $\rho$, where dilaton grows, correspond to smaller values of $\theta$. The fact that $\dot\theta <0$ at the branching point implies that $\dot\rho_+>0$. As a result bifurcation can happen only after the AdS brane has passed $\rho=\rho_m$.

Using \eqref{dotr1}, the junction condition at the dS-AdS domain wall can be simplified to a one-dimensional motion in a potential
\be\label{energy}
\dot\theta^2 + V(\theta) =0,
\ee
where
\be
V(\theta) = \left(\frac{(B\cos\theta+B_0)^2-\kappa^2 - B^2 \sin^2\theta}{2 \kappa B\sin\theta}\right)^2 -1.
\ee
The Euclidean bounce solution will be in the region $-1\leq V(\theta)\leq 0$, which corresponds to 
\be
\sqrt{\kappa^2 + B^2\sin^2\theta}\leq B\cos\theta + B_0 \leq \kappa + B\sin\theta.
\ee
To ensure that these condition are satisfied for a finite range of $\theta$, we impose
\be\label{bounds}
\sqrt{2(\kappa^2+B^2)}< B_0< \sqrt{2}B+ \kappa.
\ee
See figure \ref{fig:V} for a sketch of the potential.
\begin{figure}[t]
\centering
\includegraphics[scale =0.8]{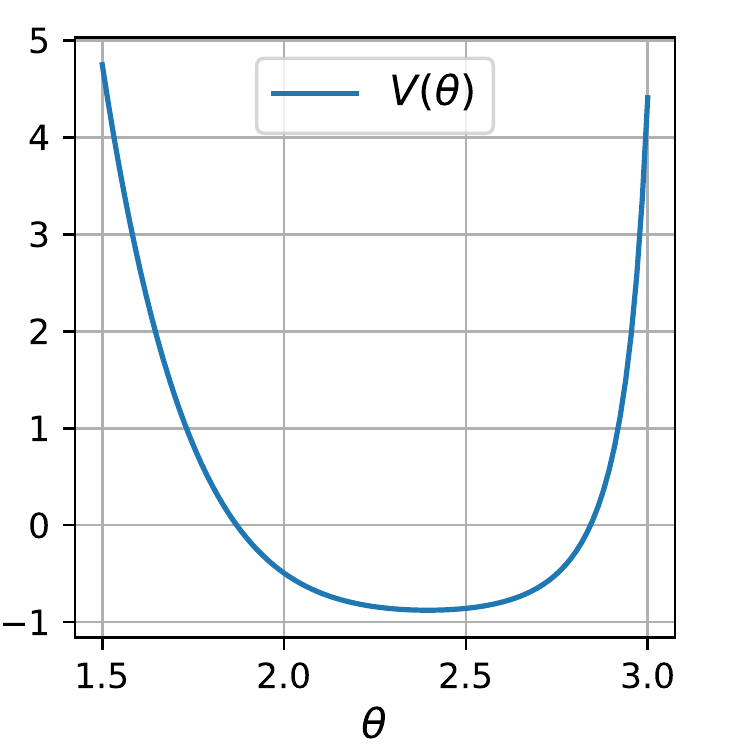} 
\caption{\small{The Euclidean potential for the dS-AdS domain wall. The parameter choices are $\kappa=0.35 B$ and $B_0=1.6 B$.}}
\label{fig:V}
\end{figure}
Wick rotation to Lorentzian signature has to be done at one of the turning points:
\be\label{th-}
\theta_- = \arccos\left(\frac{\kappa - B_0}{\sqrt{2} B}\right) - \frac{\pi}{4},
\ee
\be
\theta_+ = \frac{7\pi}{4}- \arccos\left(\frac{\kappa - B_0}{\sqrt{2} B}\right).
\ee
The potential for the Lorentzian motion is $-V(\theta)$. Hence, if the rotation is done at the larger root $\theta_+$, the walls are guaranteed to fall through the black hole horizon ($\theta = \pi$) and the cosmological region of the SdS geometry will inflate. The starting point of the bounce solution $\theta_1$ is obtained from \eqref{alpha} and \eqref{alpha1}. Using \eqref{dotr1} this can be written as
\be
F(\theta_1) = \frac{\sinh \rho_m}{\sinh \rho_1} \sqrt{1-\dot\rho_1^2}
+\left(1- \frac{B\sin\theta_1}{B\cos\theta_1+B_0}\sqrt{1-\frac{\sinh^2\rho_m}{\sinh^2 \rho_1}}\right)\dot\theta_1 =0,
\ee
where $\dot\theta_1 = -\sqrt{-V(\theta_1)}$, and $\rho_1$ and $\dot \rho_1$ are implicitly functions of $\theta_1$ and they also depend on $A$ and $\kappa_0$. It would be convenient to eliminate $A$ and $\kappa_0$, and instead consider $\rho_1$ and $\rho_m$ as independent parameters. It is then easy to see that this equation has a solution when $\rho_m\ll \rho_1$, and that this solution is close to $\theta_-$. In the limit $\rho_m\ll \rho_1$
\be
F(\theta)\simeq \frac{\sinh \rho_m}{\sinh \rho_1}\sqrt{1-\dot \rho_1^2}
- \frac{B\cos\theta +B_0 - B\sin\theta}{B\cos\theta+B_0}\sqrt{-V(\theta)}.
\ee
We have 
\be
F(\theta) \simeq\left\{\begin{array}{cc}\frac{\sinh \rho_m}{\sinh \rho_1} >0,&\qquad \theta\to \theta_-\\[10pt]
- \frac{B\cos\theta +B_0 - B\sin\theta}{B\cos\theta+B_0}\sqrt{-V(\theta)}<0,&\qquad \theta-\theta_- \gg \frac{\sinh^2\rho_m}{\sinh^2\rho_1}\end{array}\right.
\ee
so there is a solution. At this solution
\be
\dot\theta_1 \simeq -\frac{\sinh \rho_m}{\sinh \rho_1}\frac{B\cos\theta +B_0}{\kappa},
\ee
and $\theta_1-\theta_- \simeq -\dot\theta_1^2/V'(\theta_-) = \O(\sinh \rho_m/\sinh \rho_1)^2\ll 1$. 

The solution we are interested in first bounces at $\theta_-$, right after the bifurcation point, then at $\theta_+$ and for a second time at $\theta_-$, right before meeting the second bifurcation point. The point of time-reflection symmetry is $\theta = \theta_+$. In order for the interior to be free from conical singularity, we have to make sure that the bounce is completed over $\Delta\vphi_{\rm dS} =\pi$. In the above approximation,
\be
\Delta\vphi_{\rm dS} \simeq 2 \int_{\theta_-}^{\theta_+} d\theta \frac{\sqrt{1+V(\theta)}}{\sin\theta\sqrt{-V(\theta)}}=\pi.
\ee
This imposes one constraint on $B,B_0,\kappa$, which we numerically solved for and plotted in figure \ref{fig:cons}. 
\begin{figure}[t]
\centering
\includegraphics[scale =0.8]{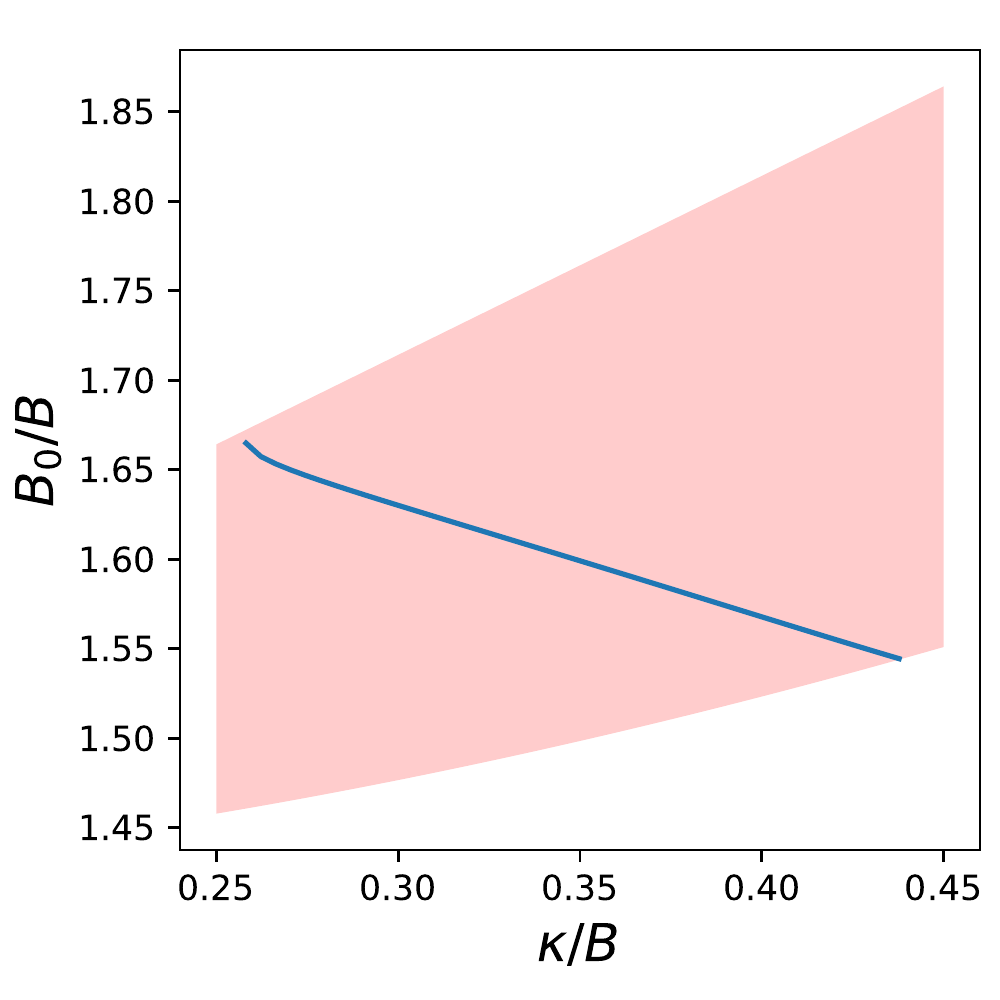} 
\caption{\small{Absence of conical singularity in the dS region imposes one constraint (blue line) on $B,B0,\kappa$. The shaded region corresponds to the range \eqref{bounds}, where the potential admits a Euclidean bounce.}}
\label{fig:cons}
\end{figure}

Finally, we should make sure that a piece of AdS boundary remains in the solution. This piece would then cross the time-reflection cut at two points and Wick rotates into the boundaries of the Lorentzian AdS$_2$ region as in figure \ref{fig:penrose}. In the full $4d$ solution they are connected to the Minkowski asymptotics. The angular size of this piece must satisfy
\be\label{vphiad}
\vphi_{b} =4\pi- 4(\Delta_1+\Delta_2 +\Delta_3)>0
\ee
where using \eqref{cosphi}
\be\label{Dphi1,2}
\Delta_1 =\arccos (\tanh \rho_m) ,\qquad \Delta_2 = \arccos \frac{\tanh \rho_m}{\tanh \rho_1}
\ee
and
\be\label{Dphi3}
\Delta_3 =\frac{B\cos\theta_- +B_0}{\sinh \rho_1} 
\int_{\theta_-}^{\theta_+}d\theta
\frac{\sqrt{1+\left(\frac{B\sin\theta}{B\cos\theta+B_0}\right)^2 V(\theta)}}{(B\cos\theta+B_0)\sqrt{-V(\theta)}}.
\ee
All $\Delta$'s can be made much less than $1$ by choosing $\rho_1\gg \rho_m \gg 1$. 
\section{Speculations}\label{sec:con}
Relative probabilities are obtained by comparing the norm of various branches of the wavefunction. Here the comparison is between the two-sided RN geometry containing a brane in the middle, and the same geometry but with the brane replaced by an expanding dS bubble. A common approach is to apply saddle-point approximation to the Euclidean gravity path integral to calculate the square of the norms (as in the standard example of Coleman-De Luccia (CDL) tunneling \cite{Coleman}): 
\be
\frac{p_{\rm dS}}{p_{\rm brane}}  \sim e^{S_0-S_b}
\ee
where $S_0$ is the Euclidean action for the freely propagating brane and $S_b$ is the bounce action. This estimate ignores the prefactor, which in particular includes the square of the coupling between the brane and domain-walls (see the end of section \ref{sec:decay}).

It is not possible to calculate $S_0$ and $S_b$ separately without specifying the embedding of the $2d$ solution in a $4d$ Euclidean solution. There is a UV ambiguity in the on-shell JT action \eqref{JT}, which contains a piece proportional to the boundary length, and hence dependent on the cutoff $\rho_c$:
\be
-2\phi_b\oint_{\d \ad} d\ell \ k \simeq 2\phi_b\ell_c +A \vphi_{b}-4\kappa_0.
\ee
Here $\ell_c =  \vphi_{b}\sinh \rho_c$, and $\vphi_{b}$ is the angular size given by \eqref{vphiad} if there is a dS bubble, and the same expression with $\Delta_2=\Delta_3=0$ if there is none. We also used \eqref{phiads},\eqref{dotr},\eqref{alb} and assumed $\phi_b\simeq A\cosh\rho_c\gg B_0$, and $\rho_m\ll \rho_c$ to get the last two terms.

It makes sense to compare solutions that have the same $\ell_c$, which is expected to be determined by the UV. For instance, if the black holes are pair produced in a magnetic field, as in \cite{Garfinkle}, the boundary length is fixed to $\sim M/Q B$ and the UV action is $\sim M \ell_c$. See \cite{Horowitz} for other formation scenarios that fix $\ell_c$. Given the change in $\vphi_{b}$ when a dS bubble nucleates, equal $\ell_c$ means different $A$ because $\ell_c \simeq \frac{\vphi_{b} \phi_b}{A}$. Equivalently, it means slightly different $M_e$ as follows from \eqref{A}. With this choice, we obtain a UV insensitive difference
\be
S_b-S_0 = \O(\phi r_0^2/G)\ll S_{BH},
\ee
where $S_{BH}$ is the Bekenstein-Hawking entropy of the black hole. In particular, in the limit $\rho_1\gg \rho_m\gg 1$, the difference is dominated by the action of the dS-AdS domain wall:
\be\label{DS}
S_b - S_0 \simeq \frac{2\kappa}{G} \int_{\theta_-}^{\theta_+} d\ell,\qquad \text{when}\quad \rho_1\gg \rho_m\gg 1.
\ee
For the choice of parameters in the plot \ref{fig:V}, we get $S_b -S_0 \simeq 1.0\frac{B r_0^2}{G}$. In the limit $A,B\to 0$, even though our estimate gives $p_{\rm dS}/p_{\rm brane}\sim 1$, the probability of forming the original RN wormhole is expected to vanish since $\ell_c\to \infty$. To conclude, the uptunnelling probability is exponentially suppressed, but it is much larger than $e^{-S_{BH}}$.

These arguments suggest that the nucleation of a false vacuum bubble is possible, assuming all ingredients are carefully prepared. However, one of the ingredients, namely the smallness of the wall tension $\kappa=8\pi G\sigma r_0 \ll 1$, is completely out of control of the experimentalist. Moreover, if the tension is too low the false vacuum region will entirely collapse (e.g. via percolating CDL bubbles). To prevent the latter, we need the critical size of the CDL bubble
\be
r_{\rm CDL} \sim \frac{\sigma}{\Lambda}
\ee
to be sufficiently big compared to the CC scale, and, to ensure the former, we need it to be small compared to $r_0$:
\be
\Lambda^{-1/4}\ll r_{\rm CDL} \ll (G\Lambda)^{-1/2}.
\ee
Hence, we need the CDL down-tunneling to be microscopic and unlikely. 

Even though this is a mild requirement, the full setup might seem exotic, and impractical for creating a Universe in the lab since it requires constructing the topologically nontrivial RN wormhole with a brane in the throat.

However, there are reasons to believe that it is not completely fictitious. It has been argued in \cite{Maldacena_traversable} that a traversable magnetic wormhole can be constructed in a QED-like model. (See \cite{Qi,Fu_wormhole,Maldacena_realtime} for related work and comments on the formation time.) This wormhole is kept open using the negative Casimir energy. Starting from there, it is a trivial task to get a non-traversable RN wormhole (which is all that we need here) by adding mass to the system.\footnote{In the first version of this article, we speculated about the possibility of a brief period of causal communication with the inflating region. However, this requires violating achronal null energy condition, and cannot be achieved using the Casimir energy. We thank Douglas Stanford for pointing this out.}

Lastly, our effective treatment of branes and domain walls guarantees that gravity by itself does not forbid formation of the false vacuum bubble. If the Maxwell theory in the Minkowski vacuum emerges from Higgsing a non-abelian gauge theory inside the false vacuum, the discharge of the magnetic field lines is also automatic and electromagnetism does not forbid the process either. Given that in non-gravitational theories false vacua are probed in scattering processes, it is not unimaginable that our toy model can indeed be an approximation to a scattering taking place inside a well-prepared wormhole. 
\section*{Acknowledgments}
We thank Sergei Dubovsky, Andrei Gruzinov, Kyriakos Papadodimas, Veronica Pasquarella, Shahin Sheikh-Jabbari, Eva Silverstein, Douglas Stanford, Giovanni Villadoro, and Zhenbin Yang for stimulating discussions. This work was partially supported by the Simons Foundation Origins of the Universe program (Modern Inflationary Cosmology collaboration).
\appendix
\bibliography{bibup}

\end{document}